\documentclass[letterpaper, 10 pt, conference]{ieeeconf}  

\IEEEoverridecommandlockouts                              

\usepackage{amsfonts}
\usepackage{xcolor}
\usepackage{graphicx} 
\usepackage{epsfig} 
\usepackage{subcaption}
\usepackage{amsmath} 
\usepackage{siunitx}
\usepackage{multicol}

\makeatletter

\newcommand{\Rmnum}[1]{\expandafter\@slowromancap\romannumeral #1@}
\makeatother

\usepackage{booktabs}
\newcommand{\ra}[1]{\renewcommand{\arraystretch}{#1}}
\usepackage{mathtools}
\newtheorem{remark}{Remark}
\usepackage{cite}
\usepackage{balance}

\title{\LARGE \bf
Integrating Electrochemical Modeling with Machine Learning for Lithium-Ion Batteries
}

\author{Hao Tu{$^1$}, Scott Moura{$^2$}, Huazhen Fang{$^1$}
\thanks{{$^1$}Hao Tu and Huazhen Fang are with the Department of Mechanical Engineering, University of Kansas, Lawrence, KS 66045, USA.
        {\tt\small tuhao@ku.edu, fang@ku.edu}}%
\thanks{{$^2$}Scott Moura is with the Department of Civil and Environmental Engineering, University of California, Berkeley, CA 94720, USA.
        {\tt\small smoura@berkeley.edu}}%
}

\begin{document}

\maketitle
\thispagestyle{empty}
\pagestyle{empty}

\begin{abstract}

Mathematical modeling of lithium-ion batteries (LiBs) is a central challenge in advanced battery management. This paper presents a new approach to integrate a physics-based model with machine learning to achieve high-precision modeling for LiBs. This approach uniquely proposes to inform the machine learning model of the dynamic state of the physical model, enabling a deep integration between physics and machine learning. We propose two hybrid physics-machine learning models based on the approach, which blend a single particle model with thermal dynamics (SPMT) with a feedforward neural network (FNN) to perform physics-informed learning of a LiB's dynamic behavior. The proposed models are relatively parsimonious in structure and can provide considerable predictive accuracy even at high C-rates, as shown by extensive simulations.


\end{abstract}

\section{Introduction}


Lithium-ion batteries (LiBs) have found fast-growing use in many industrial sectors, such as electric vehicles, smart grid, renewable energy and consumer electronics. This trend has driven a surge of research on dynamic modeling for LiBs, which plays a foundational role in advanced management of LiB systems~\cite{5466167}. Various practical applications strongly demand LiB models that can offer both high predictive accuracy and computational efficiency with a broad range of C-rates. This need, however, has not been fully met yet.

The equivalent circuit models (ECMs) are widely used in the battery engineering sector. They leverage electrical circuits to simulate a LiB's current/voltage dynamics. Their structural and mathematical simplicity make them well conducive to control and estimation design. However, with the same reason, ECMs offer relatively limited accuracy, even though recent studies have led to some more accurate designs~\cite{Ning:TCST:2020}. Electrochemical modeling is another important means to describe the dynamics of LiBs, which uses electrochemical principles to comprehensively characterize the electrochemical reactions, Li-ion diffusion and concentration changes in the electrode/electrolyte as well as various associated processes during charging/discharging of LiBs. Compared to ECMs, they can reproduce the current/voltage responses with much better accuracy. A well-known electrochemical model is the so-called Doyle-Fuller-Newman (DFN) model, which is broadly considered reliable and precise enough for almost all LiB management scenarios~\cite{Doyle_1993}. The accuracy yet comes with enormous computational complexity. This hence has motivated an incessant search for streamlined electrochemical models. The single particle model (SPM) is the most parsimonious one, which represents each electrode as a spherical particle and delineates Li-ion intercalation and diffusion in the particles~\cite{Santhanagopalan:JPS:2006}. The SPM has spurred a wide range of improved models for higher accuracy under different conditions. They usually supplement it with characterization of thermal behavior~\cite{Guo:JES:2011, Tanim:DSMC:2015}, electrolyte dynamics~\cite{Rahimian:JPS:2013, Moura:TCST:2016, Han:JPS:2015, Li:JES:2017}, and stress buildup~\cite{Li:JES:2017}. Another important line of research lies in applying numerical model order reduction methods to the DFN or other electrochemical models, with the aim of enabling computationally fast implementation~\cite{XIA20172169, 5400816, Rahn, Annaswamy, Fathy}.

Both ECMs and electrochemical models can be viewed as physics-based models as they build upon physical principles. Meanwhile, the growing data abundance for today's LiB systems due to the ubiquitous onboard sensing makes it appealing to identify models simply from measurement data. Machine learning (ML) holds much promise here, with its success in various data-driven modeling tasks. A close inspection shows that these two ways of modeling are constructively complementary. Physics-based models can offer physical interpretations of LiBs' dynamic behaviors and extrapolate to any  charging/discharging scenarios meeting the model assumptions. However, they either require much computation, as in the case of the DFN, or have inadequate accuracy when the model assumptions are not satisfied---for instance, ECMs and the SPM are designed for low/medium C-rates and can poorly predict LiBs' dynamics at high C-rates. ML-based modeling extracts black-box approximations from data in a convenient and efficient manner and requires only fixed computational costs once a model is trained. Yet, in practical applications, its performance can be constrained by the data informativeness and may face the pitfall of producing physically inconsistent results. Therefore, hybrid physics-ML modeling has emerged recently, with the aim of combining the respective merits and overcoming the challenges of each way. The study in~\cite{Refai:DSCC:2011} couples a one-dimensional electrochemical model with different kinds of neural networks (NNs). In~\cite{Park:ACC:2017}, recurrent NNs are used to learn the residuals between a LiB's terminal voltage and the SPM's output voltage. In~\cite{Feng:JPS:2020}, a simplified SPM and a lumped thermal model are combined with a neural network in series to predict the terminal voltage. While these studies offer promising results, the hybrid modeling of LiBs is still under-explored, without living up to its potential.  Note that a related line of research is to apply ML, e.g., NNs and Gaussian processes, to estimate the state-of-charge, state-of-health and temperature of LiBs~\cite{Chen:TCST:2018, Sahinoglu:TIE:2018, RICHARDSON2019320}, which yet is beyond the scope of this paper.

In the hybrid models surveyed above, the input of an NN includes the current and output voltage of the physical model, and the output of the NN is the residual or actual terminal voltage. However, the mapping from the NN's input to its output would not be one-on-one physically, due to the dynamics inside LiBs. The NN, and  consequently the hybrid model, will hence be limited in the predictive capability, even if achieving satisfactory training accuracy. To overcome this key limitation, this work contributes a new perspective into hybrid modeling: {\em the NN should be additionally informed of the internal state of the physical model}. This will allow the NN to be aware of the ongoing dynamics of the physical model and thus learn more effectively what is missed from the physical model in comparison to the measurement data. This perspective leads us to develop the following specific contributions.
\begin{itemize}

\item We develop two new hybrid physics-ML models for LiBs which integrate the SPMT with the FNNs. The first model, named HYBRID-I, leverages an NN to capture the residuals of SPMT, and the second, named HYBRID-II, uses an NN to predict the voltage based on the SPMT. Different from the literature, both of them feed the state-of-charge (SoC) information of the SPMT to the NN as an additional critical input.

\item We evaluate the two hybrid models by extensive simulations. They both demonstrate high predictive accuracy when applied to testing datasets and are found applicable to a broad range of C-rates as high as 10 C. The use of FNNs also make the architectures of the two models relatively parsimonious.  

\end{itemize}
With the exceptional accuracy, the proposed hybrid models can find prospective use in various energy storage   applications, especially those that require high-power charging/discharging.

This paper is organized as follows. Section II discusses the construction of the proposed hybrid models. Section III presents an overview of the SPMT and FNN models. Section IV contains evaluation settings and simulation results. Finally, section V concludes our work with remarks and future research.

\section{Hybrid Physics-ML Modeling for LiBs}\label{Hybrid-Modeling}

This section constructs hybrid models that blend the SPMT with an FNN to describe the current/voltage dynamics of LiBs. The aim is leveraging the SPMT to create a physics-informed FNN, thus endowing the hybrid models with high predictive accuracy and ability across a broad C-rate range. Next, we begin with a brief introduction to the SPMT and FNNs and then present the hybrid model design. A further description of the SPMT and FNN is offered in Section~\ref{SPMT-FNN}.

\begin{figure}
    \centering
    \includegraphics[width = 0.5\textwidth,trim={6.5cm 2cm 5.7cm 5cm},clip]{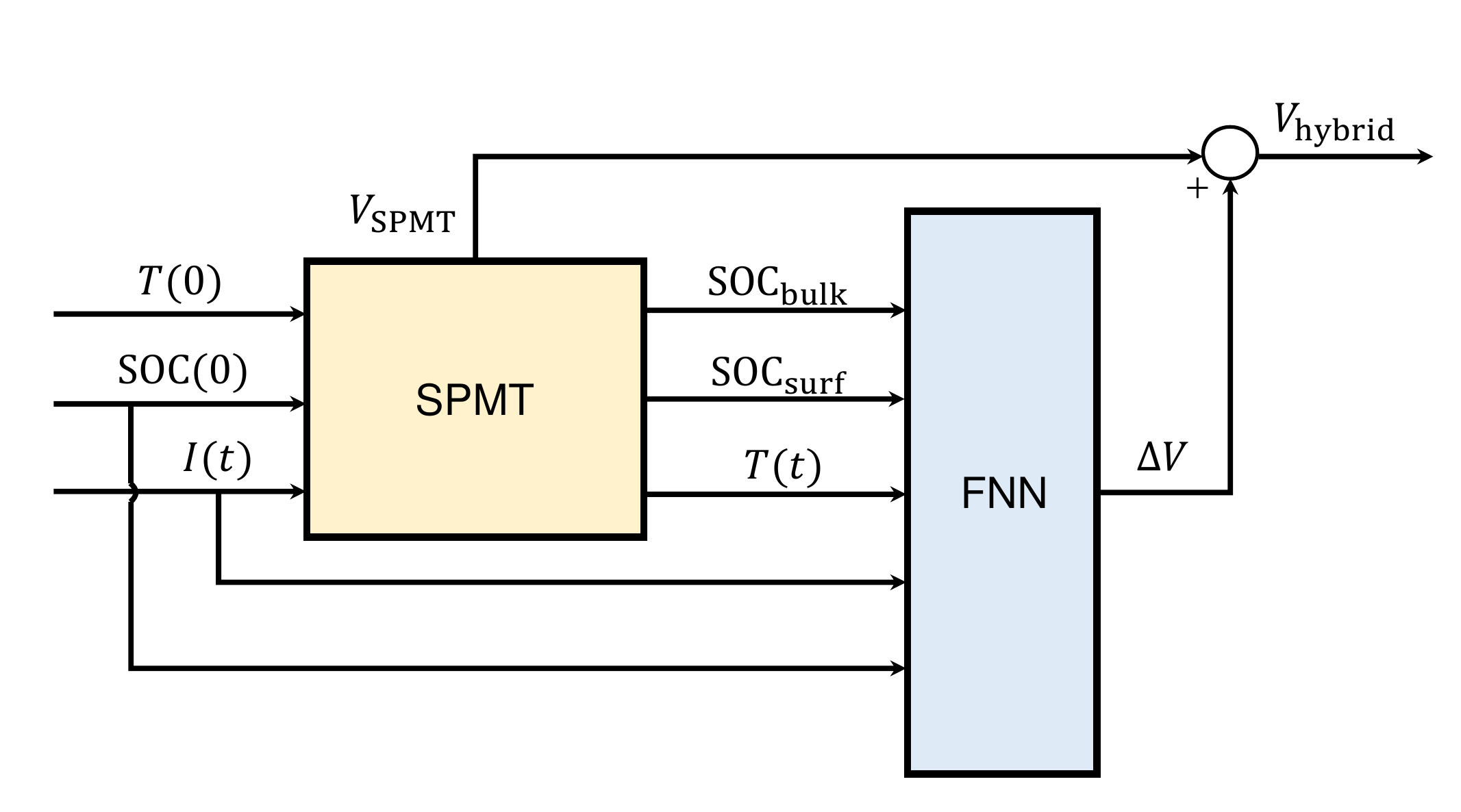}
    \caption{Block diagram of HYBRID-\Rmnum{1}}
    \label{fig: HYBRID1}
\end{figure}

\begin{figure}
    \centering
    \includegraphics[width = 0.5\textwidth,trim={4.5cm 4cm 8.5cm 6cm},clip]{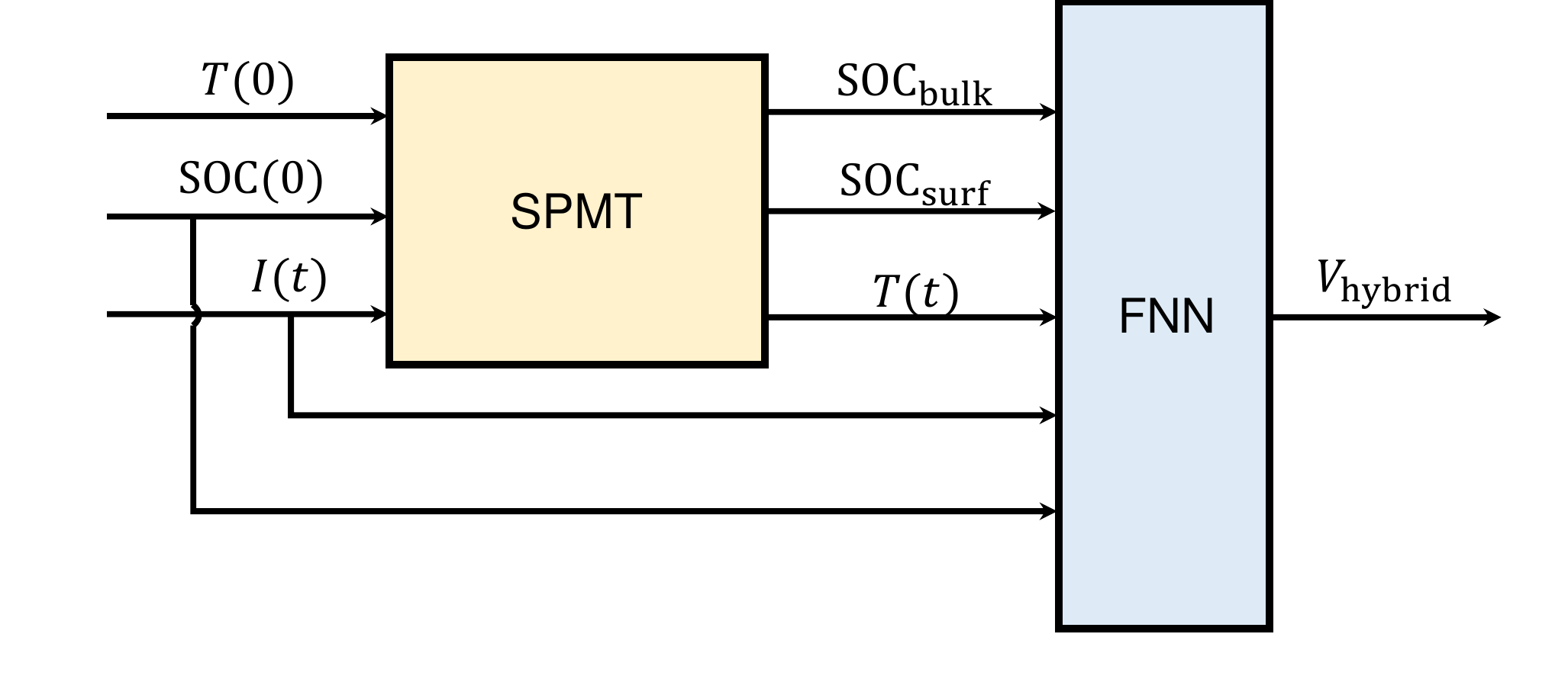}
    \caption{Block diagram of HYBRID-\Rmnum{2}}
    \label{fig: HYBRID2}
\end{figure}

The SPMT is an improved version of the SPM, which contains an extra bulk thermal model to describe a LiB's thermal behavior~\cite{Bernardi_1985}. Its dynamics hence mainly includes the lithium-ion diffusion in the electrodes, electrochemical reaction kinetics on the surface of the electrodes, and lumped thermodynamics. When a charging/discharging current $I(t)$ is applied to the SPMT, the lithium-ion concentrations in the electrodes, $c_s^\pm$, will change due to the reaction and transport, and so will the lumped temperature $T(t)$; consequently, the SPMT's output voltage, $V_{\mathrm{SPMT}}$, will evolve through time. The SPMT is more accurate than the SPM but still valid for only low to medium C-rates (up to 1C)~\cite{Guo:JES:2011}.

FNNs are  an important class of ML methods designed to approximate complex functions. An FNN is composed of multiple layers of different interconnected functions, analogous to a brain comprising many interconnected neurons. Its network structure contains no cycle or feedback connections, making FNNs the simplest type of NNs and  easier to train and implement. The theoretical performance of FNNs is guaranteed by the universal approximation theorem, which generally states that a continuous vector-valued function in the real space can be approximated with arbitrary accuracy by an FNN~\cite{HORNIK1989359}. 

We develop our first hybrid model, HYBRID-I, based on residual modeling. Its structure is depicted in Fig.~\ref{fig: HYBRID1}. As aforementioned, the SPMT as a reduced electrochemical model  is limited to low to medium C-rates   and would give poor prediction  at high C-rates.  An FNN is thus used to learn the residual errors of the SPMT relative to the measurements. The FNN's output then is $\Delta V = V - V_{\mathrm{SPMT}}$, where $V$ is the actual voltage. The input selection is critical, as it decides the information   that the FNN can exploit. A primary criterion is to enable the FNN to utilize the SPMT-based physics both sufficiently and economically. 

Here, we designate the input to include the current $I(t)$, temperature $T(t)$, initial SoC $\mathrm{SoC(0)}$, and bulk and surface SoC in the negative electrode, denoted as $\mathrm{SoC}_{\mathrm{bulk}}$ and $\mathrm{SoC}_{\mathrm{surf}}$, respectively. Note that $\mathrm{SoC}_{\mathrm{bulk}}$ and $\mathrm{SoC}_{\mathrm{surf}}$ offer an aggregated yet efficient representation of the SPMT's state, so feeding them into the FNN will provide an awareness of the SPMT's overall internal condition, in the benefit of residual learning and prediction. 

The second hybrid model, HYBRID-II, is shown in Fig.~\ref{fig: HYBRID2}. It connects the SPMT and an FNN in cascade to learn the terminal voltage directly. The FNN uses the same set of variables from the SPMT as in HYBRID-I, but its output tries to emulate $V$. By design, this structure also uses the SPMT-based physics to ensure the FNN's effectiveness of learning.

\begin{remark}
The pivotal difference of the above hybrid modeling design from the literature is that information about the physical model's state is fed as part of the input to the ML model. This integration provides a closer physics-ML integration than the means described in the literature. It particularly allows the ML to associate the physical model's inherent  internal dynamics  with the measurement data, thus greatly enhancing the predictive accuracy. This design is the first of its kind to our knowledge.
\end{remark}

\begin{remark}
The choice is non-unique for the variables used to represent the SPMT's state and feed to the FNN. For instance, an expedient way is to just use the full state of the SPMT. This, however, will cause extremely high training and computational costs. We find out that just some simple, aggregated state variables will suffice. After much trial-and-error search, we identify that 
$\mathrm{SoC}_{\mathrm{bulk}}$ and $\mathrm{SoC}_{\mathrm{surf}}$ offer a favorable choice in terms of both computational efficiency and prediction performance.
\end{remark}

\section{Overview of SPMT and FNNs}\label{SPMT-FNN}

Following Section~\ref{Hybrid-Modeling}, this section provides an overview of the SPMT and FNNs. 
 
\subsection{The SPMT model}
Developed in~\cite{Guo:JES:2011}, the SPMT model couples the SPM model with a bulk thermal model to predict the electrochemical and thermal behaviors simultaneously. 

The SPM simplifies each electrode of a LiB cell as a spherical particle and neglects the electrolyte dynamics. The transport of the lithium ions inside a  particle is governed by the Fick's diffusion law in the spherical coordinates:
\begin{align*}
    \frac{\partial c_s^\pm}{\partial t}(r,t) = \frac{1}{r^2}\frac{\partial}{\partial r}\left[D_s^\pm r^2 \frac{\partial c_s^\pm}{\partial r}(r,t)\right] ,
\end{align*}
where $c_s^\pm(r,t)$ is the solid-phase lithium-ion concentration of positive ($+$) or negative ($-$) electrode, and $D_s^\pm$ is solid-phase diffusion coefficient. The diffusion is subject to the following boundary conditions: 
\begin{align*}
    \frac{\partial c_s^\pm}{\partial r}(0,t) = 0 ,   \ 
    \frac{\partial c_s^\pm}{\partial r}(R_s^\pm,t) = -\frac{1}{D_s^\pm} j_n^\pm ,
\end{align*}
where $R_s^\pm$ is the particle  radius and $j_n^\pm$ is the molar flux at the particle surface. Here, $j_n^\pm$ is given by
\begin{align*}
    j_n^\pm = \mp \frac{I(t)}{a_s^\pm F A L^\pm} ,
\end{align*}
where $a_s$ is the specific interfacial area, $F$ is the Faraday's constant, $A$ is an electrode's surface area, and $L$ is an electrode's thickness. Further, $j_n^\pm$ results from  the electrochemical kinetics and depends on the overpotential of the electrodes $\eta^\pm$. The relation is  characterized by the Butler-Volmer equation:
\begin{align*}
j_n^\pm = \frac{1}{F} i_0^\pm\left[ \exp\left({\frac{\alpha_{a}F}{RT}}\eta^{\pm}\right)-\exp\left({\frac{-\alpha_{c}F}{RT}\eta^{\pm}}\right) \right] ,
\end{align*}
where the exchange current density $i_0^\pm$ is a function of kinetic reaction rate $k^\pm$, constant electrolyte-phase lithium-ion concentration $c_e^{0}$, solid-phase lithium-ion concentration at the particle surface $c_{ss}^\pm(t)=c_s^\pm(R_s^\pm,t)$ and maximum concentration in the solid phase $c_{s,\max}^\pm$
\begin{align*}
    i_0^\pm = k^\pm (c_e^{0})^{\alpha_{a}}(c_{ss}^\pm(t))^{\alpha_{c}}({c_{s,\max}^\pm - c_{ss}^\pm(t))^{\alpha_{a}}} ,
\end{align*}

By assuming anodic and cathodic charge transfer coefficient $\alpha_{a}=\alpha_{c}=0.5$, the above indicates that $\eta^\pm$ can be expressed as
\begin{align*}
\eta^\pm = \frac{2RT}{F}\sinh^{-1}\left(\frac{F}{2i_0^\pm}j_n^\pm\right) ,
\end{align*}
The terminal voltage $V$ is  

\begin{align*}
    V(t) &= U^+(c_{ss}^+(t))-U^-(c_{ss}^-(t))+\eta^+ -\eta^- \\ & \quad  - \left(\frac{R_f^+}{a_s^+L^+}  + \frac{R_f^-}{a_s^-L^-}\right)I(t) ,
\end{align*}
where $U^+$, $U^-$ are equilibrium potential and $R_f^+$, $R_f^-$ are solid-electrolyte interphase film resistance.


The charging/discharging of LiBs is accompanied by  temperature buildup. The change in temperature can be intense at large currents and notably affects the lithium-ion diffusion and electrochemical kinetics. Characterizing and incorporating this effect will help  improve the accuracy of the SPM. Here, the temperature dependence of $D_s^\pm$ and $k^\pm$ is the most important, which is governed by the Arrhenius law:
\begin{align*}
    \psi = \psi_{\mathrm{ref}}\exp\left[\frac{E_{\psi}}{R}\left(\frac{1}{T_{\mathrm{ref}}}-\frac{1}{T(t)}\right)\right] ,
\end{align*}
where $\psi$ is $D_s^\pm$ or $k^\pm$, $T(t)$ is the lumped temperature, $R$ is universal gas constant and $E_{\psi}$ is activation energy.
Based on the energy balance principle, the change of $T(t)$ is assumed to follow
\begin{align*}
    \rho_{\mathrm{avg}} c_p \frac{dT(t)}{dt} = \dot{q}_{\mathrm{gen}} + \dot{q}_{\mathrm{conv}} ,
\end{align*}
where $\rho_{\mathrm{avg}}$ is cell bulk density, $c_p$ is lumped specific heat capacity, $\dot{q}_{\mathrm{gen}}$ denotes the heat generation rate which is contributed by ohmic heat and entropic heat and $\dot{q}_{\mathrm{conv}}$ is the convective heat removal rate with the ambience. Further, $\dot{q}_{\mathrm{gen}} $ and $\dot{q}_{\mathrm{conv}}$ are given by
\begin{align*}
\dot{q}_{\mathrm{gen}} &= I(t)[V(t)-(U^+(\Bar{c}_{s}^+(t)) - U^-(\Bar{c}_{s}^-(t)))] \\ &\quad + I(t)T(t)\frac{\partial}{\partial T}[U^+(\Bar{c}_{s}^+(t)) - U^-(\Bar{c}_{s}^-(t))] ,\\
\dot{q}_{\mathrm{conv}} &= -h_{\mathrm{cell}}(T(t)-T_{\mathrm{amb}}(t)) ,
\end{align*}
where $T_{\mathrm{amb}}$ is the ambient temperature, $h_{\mathrm{cell}}$ is the convective heat transfer coefficient and the anodic/cathodic bulk concentration $\Bar{c}_{s}^\pm(t)$ is given by:
\begin{align*}
\Bar{c}_{s}^\pm(t) = \frac{3}{(R_s^\pm)^3} \int_{0}^{R_s^\pm} r^2 c_s^\pm(r,t) dr ,
\end{align*}
We define the anodic surface SoC and bulk SoC as:
\begin{align*}
    \mathrm{SoC}_{\mathrm{surf}} = \frac{c_{ss}^-(t)}{c_{s,\max}^-},
    \ \
    \mathrm{SoC}_{\mathrm{bulk}} = \frac{\Bar{c}_{s}^-(t)}{c_{s,\max}^-}.
\end{align*}

Summarizing the foregoing  equations will lead to a complete formulation of the SPMT model. It offers good accuracy when low to medium currents are applied. Yet, its prediction performance at high C-rates is harmed by the idealistic simplification of an electrode as a particle and absence of the electrolyte dynamics. 


\subsection{The FNN Model} 

\begin{figure}
    \centering
    \includegraphics[width = 0.5\textwidth,trim={5cm 2.2cm 5cm .9cm},clip]{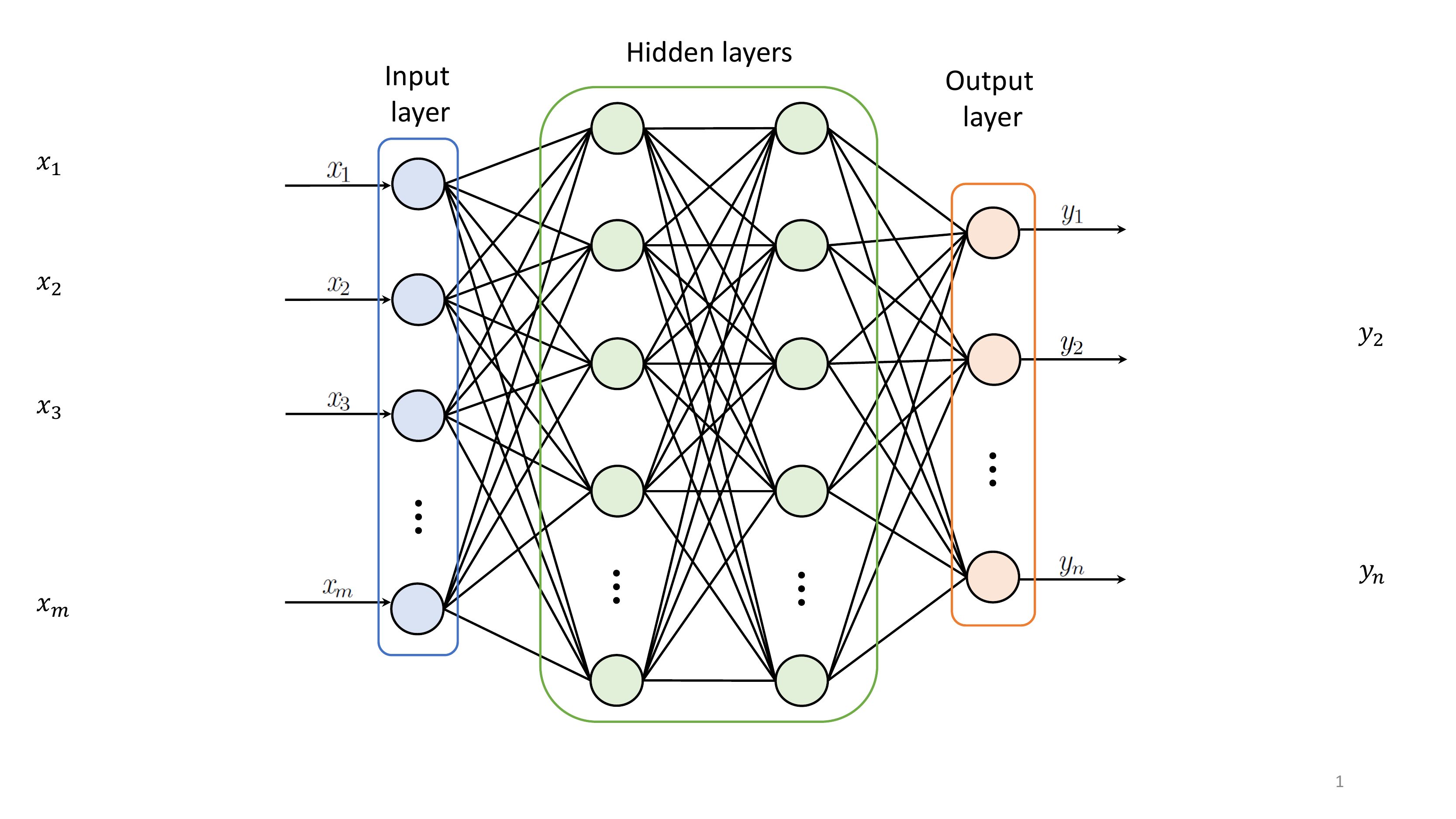}
    \caption{FNN architecture with two fully connected hidden layers}
    \label{fig:NN}
\end{figure}

FNNs are a basic yet extremely important  type of NNs, with numerous successful applications. We offer an overview of them here, which is mainly based on~\cite{Goodfellow:DeepLearningBook, Malmstrom:IFAC:2020}. 

Consider  an unknown function $g^\ast$, which is a mapping from a $m$-dimensional input $\boldsymbol  x$ to a $n$-dimensional output $\boldsymbol y$. An FNN aims to approximate it by defining  a parameterized mapping ${\boldsymbol y} = g\left(\boldsymbol x, \boldsymbol \theta \right)$ and learning the collection of parameters $\boldsymbol \theta$ from the data $\left\{ ({\boldsymbol x}_i, {\boldsymbol y}_i), i=1,2,\ldots,N \right\}$. The structure of an FNN includes   three parts interconnected in series, namely, the input layer, hidden layers, and  output layer. The input layer takes the input $\boldsymbol x$ and passes it to the first hidden layer. A hidden layer makes a nonlinear transformation of its input. For example, the first hidden layer  will transform $\boldsymbol x$ into $\phi(\boldsymbol W_0  \boldsymbol x + \boldsymbol b_0)$, where $\phi$ is a chosen nonlinear  mapping often called as activation function, $\boldsymbol W_0$ is the weight matrix, and $\boldsymbol b_0$ is a correction term. The following hidden layers  then run the same   nonlinear transformation  sequentially. The final layer is the output layer, which generates an output value to match $\boldsymbol y$. Considering an $L$-layer FNN,  it can be described in a general form: 
\begin{align*}
{\boldsymbol z}_1 &= {\boldsymbol x} ,\\
{\boldsymbol z}_{l } &= \phi \left(  {\boldsymbol W}_{l-1}  {\boldsymbol z}_{l-1} + {\boldsymbol b}_{l-1} \right), \ l =   2, 3, \ldots, L-1,\\
{\boldsymbol y} &= {\boldsymbol W}_{L-1}  {\boldsymbol z}_{L-1}  + {\boldsymbol b}_{L-1},
\end{align*}
where $ {\boldsymbol z}_{l-1}$ and  ${\boldsymbol z}_{l}$  are the input and output of the $l$-th layer, respectively. For the FNN, $\boldsymbol \theta$ is the collection of ${\boldsymbol W}_l$ and ${\boldsymbol b}_l$ for $l=1, 2, \ldots, L-1$. Note that the information flows only in the forward direction from $\boldsymbol x$ to $\boldsymbol y$  in the above network model, which is why the model is called as  feedforward NN. 

The training of the FNN is to identify $\boldsymbol \theta$ from  the measurement data $\left\{ ({\boldsymbol x}_i, {\boldsymbol y}_i)  \right\}$. A common approach is based on   maximum likelihood, which   minimizes the following cost function:
\begin{align*}
J(\boldsymbol \theta) = - \mathbb{E}_{{\boldsymbol x}, {\boldsymbol y} \sim \hat p_{\mathrm{data}} } \log p_{\mathrm{model}} \left( {\boldsymbol y} \mid {\boldsymbol x}, {\boldsymbol \theta}\right),
\end{align*}
where $\hat p_{\mathrm{data}}$ is the data-based empirical distribution of $\boldsymbol x$ and $\boldsymbol y$, and $p_{\mathrm{model}}$ is the probability distribution of $\boldsymbol y$ over the parameter space $\boldsymbol \theta$ based on the FNN model. Under some assumptions, $J(\boldsymbol \theta)$ can be reduced to a mean squared error cost: 
\begin{align*}
J(\boldsymbol \theta)  =  \frac{1}{N} \sum_{i=1}^N \left\| {\boldsymbol y}_i - g \left(  {\boldsymbol x}_i, \boldsymbol \theta \right) \right\|^2.
\end{align*}
The minimization is usually achieved using the  stochastic gradient descent algorithm.  The computation of the gradient can be complicated, especially for multi-layer FNNs,  but it can still be done efficiently and exactly by the back-propagation algorithm and its generalizations.

\section{Hybrid Model Validation}

In this section, we perform simulation to validate the effectiveness of the proposed HYBRID-I and HYBRID-II models. 

\subsection{Evaluation Setting}

The evaluation setting is summarized as follows:
\begin{itemize}

\item The DFN model, which is acknowledged as a generic and reliable electrochemical model, is used as the benchmark to assess the HYBRID-I and HYBRID-II.  

\item We use the DUALFOIL simulation package~\cite{DualFoil} to  run the DFN model representing  an LCO/graphite battery that operates between 3.1 and 4.1 V to generate   synthetic   data  as the ground truth.

\item The synthetic data are divided into the training and test datasets. The training datasets are produced by applying constant discharging currents at 0.1/0.2/1/2/4/6/8/10 C and variable currents created based on the Urban Dynamometer Driving Schedule (UDDS) and US06 at $\mathrm{SOC(0)} =  0.27, 0.52, 0.67,0.74$.   The test datasets are obtained by applying constant discharging currents at 0.5/1/3/5/7/10 C and variable currents created based on the  UDDS  and US06 at $\mathrm{SOC(0)}=0.46, 0.58, 0.70$. Here, the UDDS-based training and testing current profiles are generated by modifying the standard UDDS profile differently
so that they differ from each other notably. They are labeled as UDDS-A and UDDS-B, respectively. Besides, all variable current profiles are scaled to a maximum current of around 10 C. In all cases, the initial temperature $T(0)=T_{amb}=25^\circ$C. 

\item  Both the HYBRID-I and HYBRID-II employ a  four-layer  FNN. The FNN has two  hidden layers, each consisting of 32 nodes. The input and output of the FNN are as specified in Section \ref{Hybrid-Modeling}. The rectified linear unit (ReLu) function  is chose as the activation function for the two hidden layers, and the linear activation function chosen for the output layer. Keras, a Python-based deep learning library, is used to create, train  and implement the FNN.  Because the quantities of  the input variables vary across different orders of magnitude, the input data are pre-processed by normalization, as often recommended in the practice of NNs. 

\item We utilize  the root-mean-square error as a metric to evaluate a model's performance:
\begin{align*}
\mathrm{RMSE} = \sqrt{\frac{1}{N }\sum_{i=1}^{N } \left(V_i -V_{\mathrm{model},i}  \right)^2},
\end{align*}
where $V_i$ is the true voltage at time $i$,  $V_{\mathrm{model},i}$ is the model-based voltage prediction,   and $N $ is the total number of data values. Furthermore, a relative error reduction ($\mathrm{RER}$) is introduced to quantify the improvement of the HYBRID-I and HYBRID-II over the SPMT, which is defined as
\begin{align*}
   \mathrm{RER} = \frac{\mathrm{RMSE_{SPMT}}-\mathrm{RMSE_{Hybrid}}}{\mathrm{RMSE_{SPMT}}} \times 100\%.
\end{align*}

\end{itemize}

\subsection{Validation of the HYBRID-I and HYBRID-II Models}

\begin{figure}[h!]
      \centering
      \includegraphics[width = 0.455\textwidth,trim={.5cm 0cm .5cm .7cm},clip]{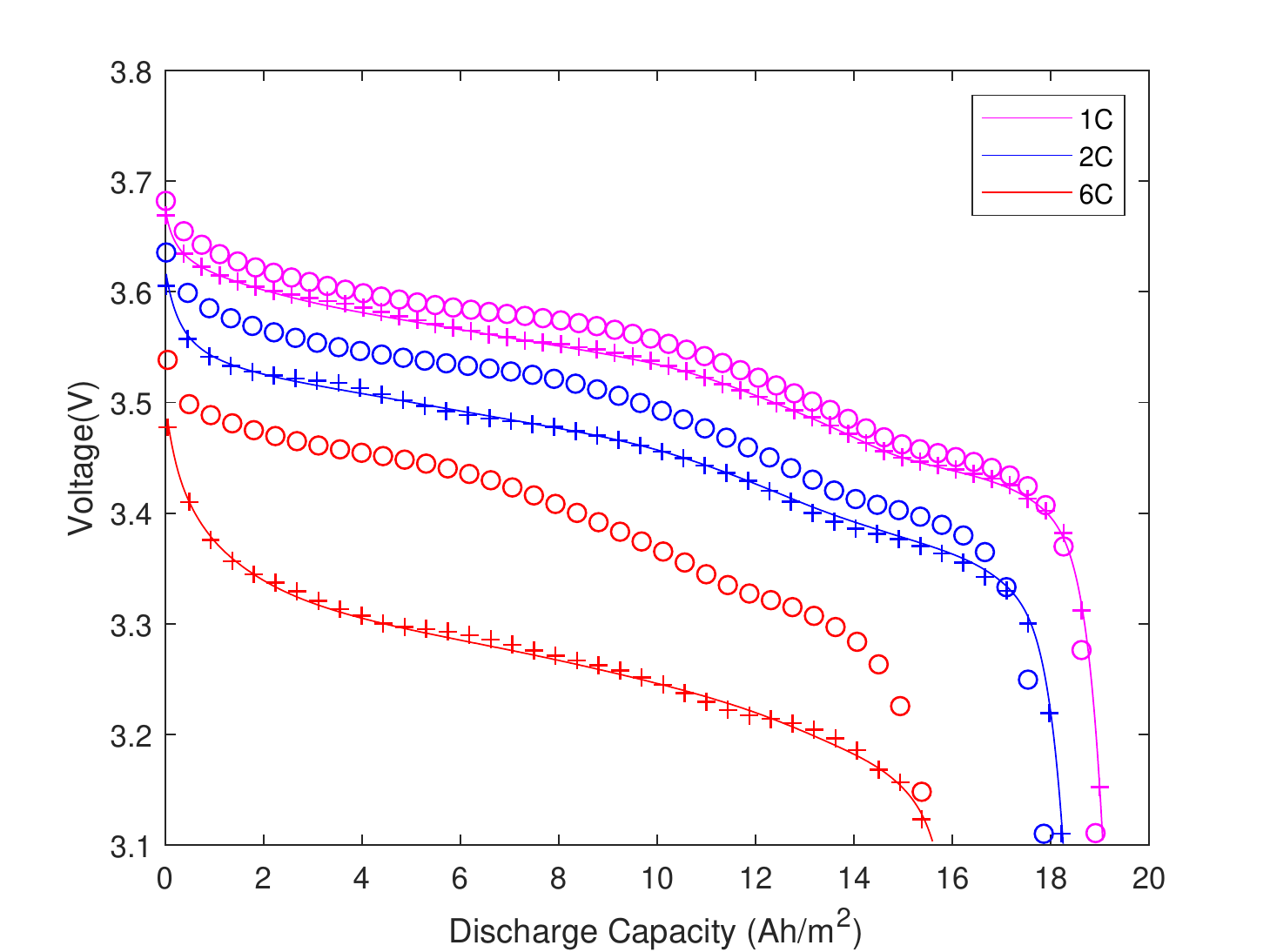}
      \caption{Training results under 1/2/6 C constant-current discharging when $\mathrm{SOC}(0) = 0.52$. Marker symbols: line ``$-$'' for DFN;  circle ``$\circ$'' for SPMT; plus ``$+$'' for HYBRID-\Rmnum{1}.}
      \label{HYBRID-I-training-constant-current}
\end{figure}

\begin{figure}[h!]
    \centering
    \includegraphics[width = 0.455\textwidth,trim={.5cm .9cm .5cm 1cm},clip]{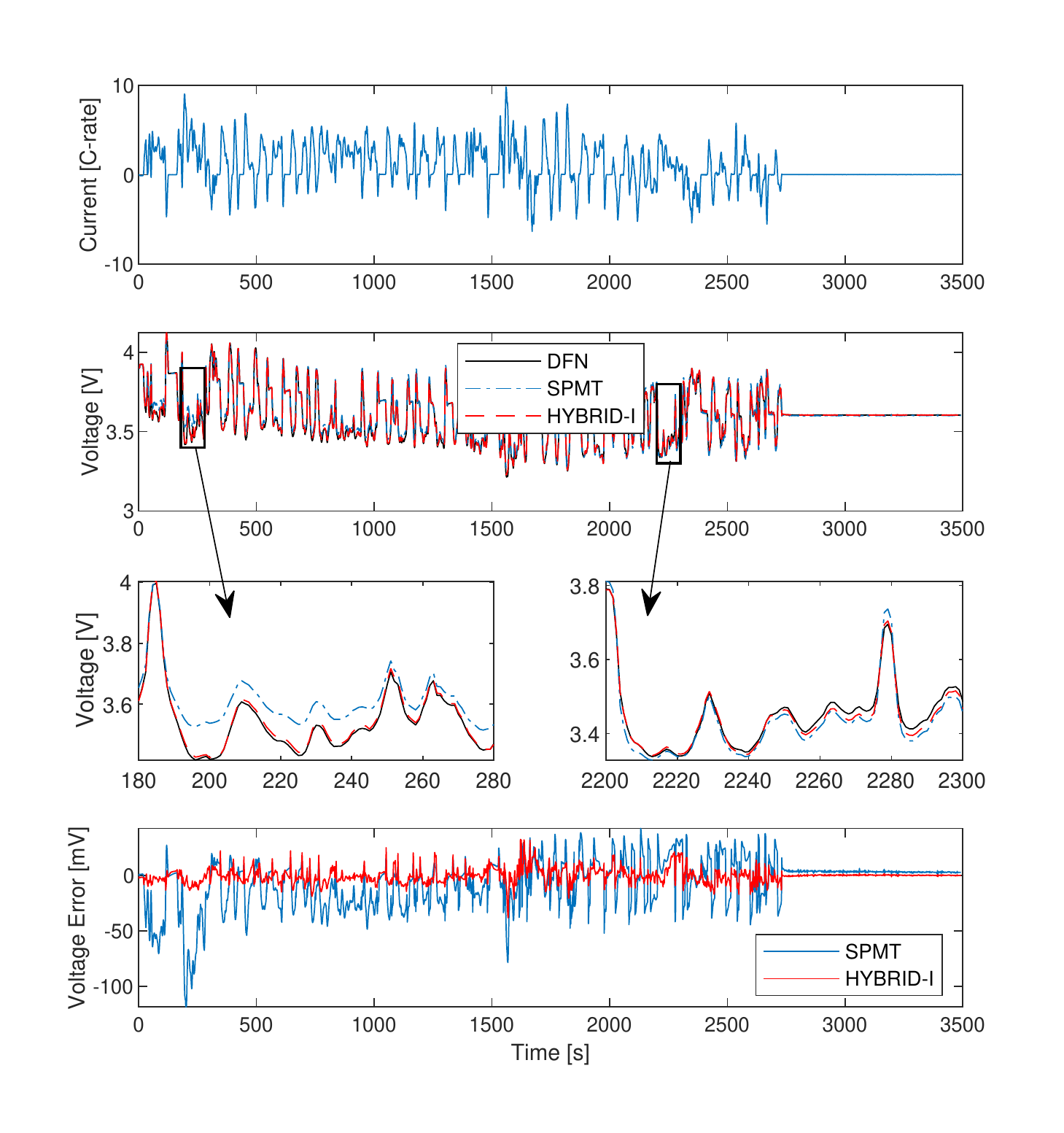}
    \caption{Training results of HYBRID-\Rmnum{1} under UDDS-A discharging with  $ \mathrm{SOC}(0) = 0.67$. }
    \label{HYBRID-I-training-UDDS-A}
\end{figure}

\begin{figure}[h!]
      \centering
      \includegraphics[width = 0.455\textwidth,trim={.5cm 0cm .5cm .7cm},clip]{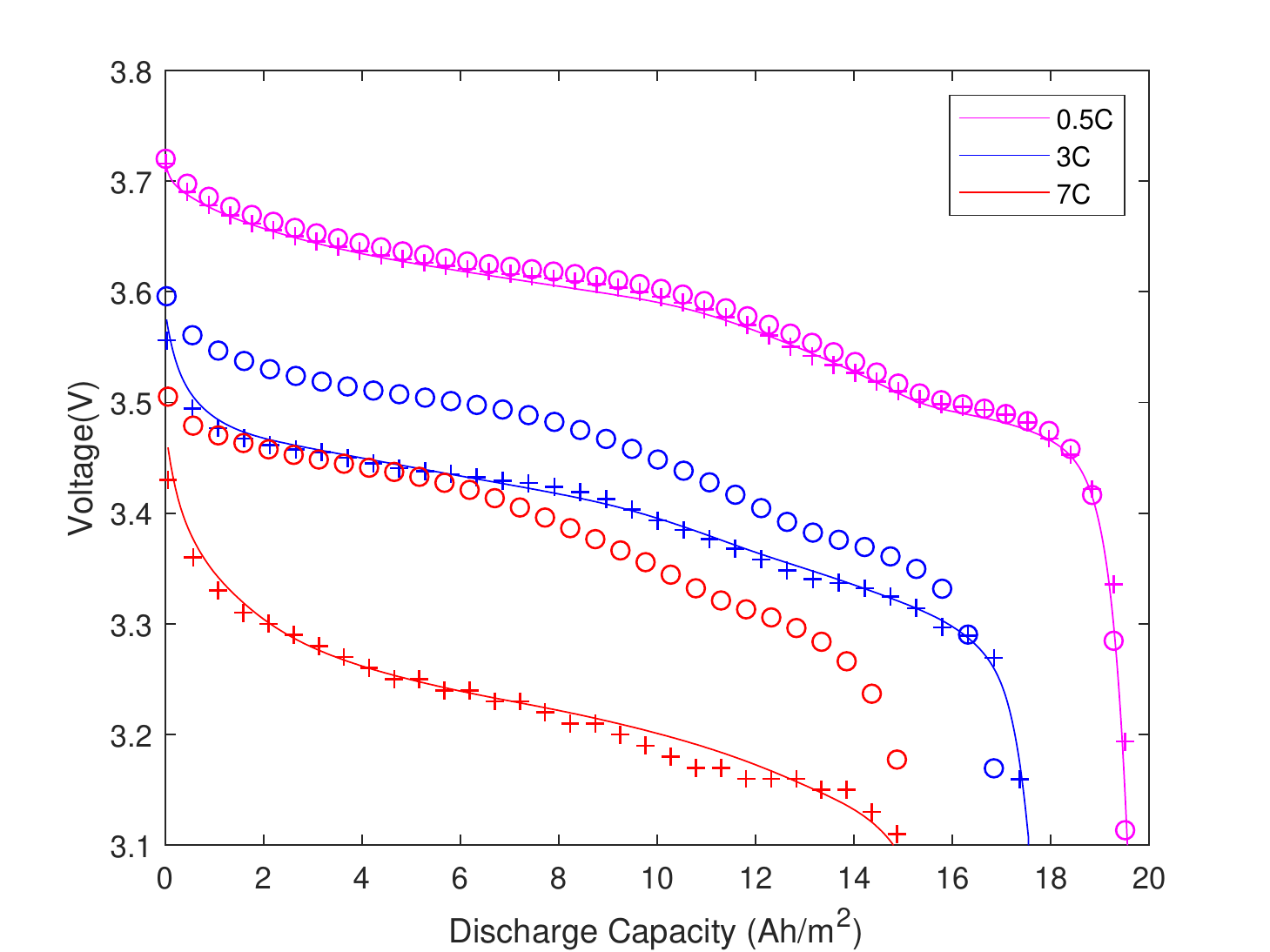}
      \caption{Testing results under 0.5/3/7 C constant-current discharging for $\mathrm{SOC}(0) = 0.58$. Marker symbols: line ``$-$'' for DFN;  circle ``$\circ$'' for SPMT; plus ``$+$'' for HYBRID-\Rmnum{1}.}
      \label{HYBRID-I-test-constant-current}
\end{figure}

\begin{figure}[h!]
    \centering
    \includegraphics[width = .455\textwidth,trim={.5cm .9cm .5cm 1cm},clip]{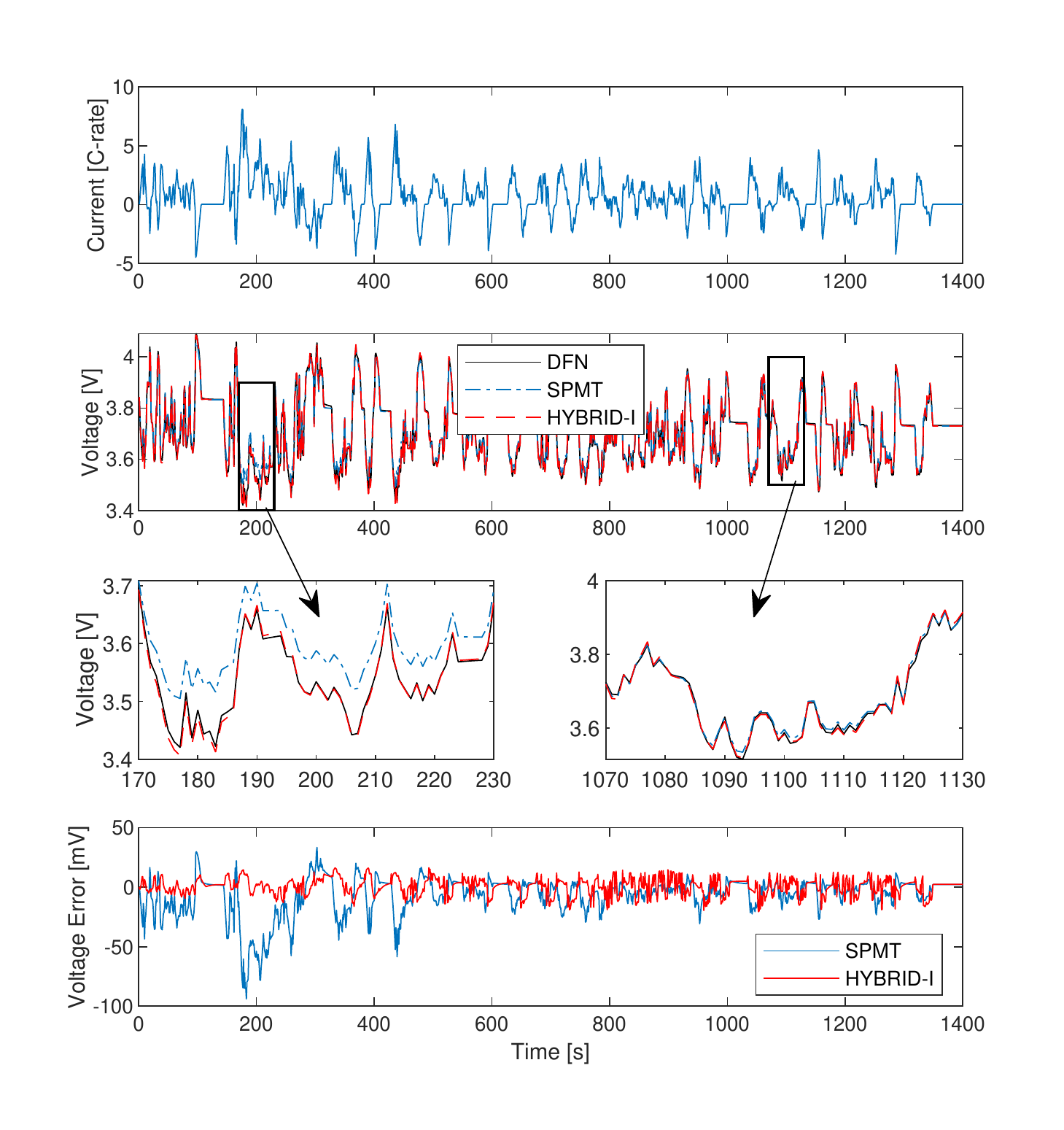}
    \caption{Testing results of HYBRID-\Rmnum{1} under UDDS-B discharging for $\mathrm{SOC}(0) = 0.58$.}
    \label{HYBRID-I-test-UDDS-B}
\end{figure}

We begin with validating the HYBRID-I model. Table~\ref{HYBRID-I-training-all-datasets} summarizes its performance over all the training datasets and compares it with the baseline SPMT. Further, Figs.~\ref{HYBRID-I-training-constant-current}-\ref{HYBRID-I-training-UDDS-A} offers a visual evaluation of the training results under constant and variable (UDDS-A) current profiles. We observe that the HYBRID-I offers remarkable accuracy  in all training scenarios. It consistently outperforms the SPMT with a considerable margin, especially at medium to very high currents.  Further, we apply the trained HYBRID-I model to the test datasets to appraise its prediction performance. 
Table~\ref{HYBRID-I-test-all-datasets} and Figs.~\ref{HYBRID-I-test-constant-current}-\ref{HYBRID-I-test-UDDS-B} demonstrate the results. As is seen, the HYBRID-I can still  retain the high accuracy in the testing phase, proving its strong predictive ability. 

\begin{table}[h]\centering
\ra{1.2}
\label{table 1}
\begin{tabular}{l | l l l}
\toprule
Input profile & RMSE (SPMT) & RMSE (HYBRID-\Rmnum{1}) & RER (\%) \\
\hline
CC-0.1C & 2.31 mV & 1.58 mV & 31.60\\

CC-0.2C & 4.37 mV & 1.44 mV & 44.95\\

CC-1C & 21.59 mV & 2.87 mV & 86.71\\

CC-2C & 56.40 mV & 4.72 mV & 91.63\\

CC-4C & 94.28 mV & 8.88 mV & 90.58\\

CC-6C & 118.86 mV & 7.40 mV & 93.77\\

CC-8C & 150.06 mV & 7.83 mV & 94.78\\

CC-10C & 191.05 mV & 10.87 mV & 94.31\\

UDDS-A & 23.32 mV & 8.60 mV & 63.12\\

US06 & 33.30 mV & 9.65 mV & 71.02\\
\bottomrule
\end{tabular}
\caption{Training performance of the HYBRID-\Rmnum{1} under different current profiles and initial SOCs, in comparison with the SPMT.}
\label{HYBRID-I-training-all-datasets}
\end{table}

\begin{table}[h!]
\centering
\ra{1.2}
\label{table}
\begin{tabular}{l | l l l}
\toprule
Input profile & RMSE (SPMT) & RMSE (HYBRID-\Rmnum{1}) & RER (\%) \\
\hline
CC-0.5C & 10.92 mV & 7.82 mV & 28.39\\

CC-1C & 19.15 mV & 5.05 mV & 73.63\\


CC-3C & 76.37 mV & 9.19 mV & 87.97\\


CC-5C & 101.85 mV & 7.19 mV & 92.94\\


CC-7C & 137.85 mV & 8.77 mV & 93.64\\


CC-10C & 207.29 mV & 10.67 mV & 94.85\\


UDDS-B & 17.59 mV & 6.75 mV & 61.63\\

US06 & 35.05 mV & 9.12 mV & 73.98\\
\bottomrule
\end{tabular}
\caption{Testing performance of the HYBRID-\Rmnum{1} over the test datasets, in comparison with the SPMT.}
\label{HYBRID-I-test-all-datasets}
\end{table}

\begin{figure}[h!]
    \centering
    \includegraphics[width = .455\textwidth,trim={.5cm .9cm .5cm 1cm},clip]{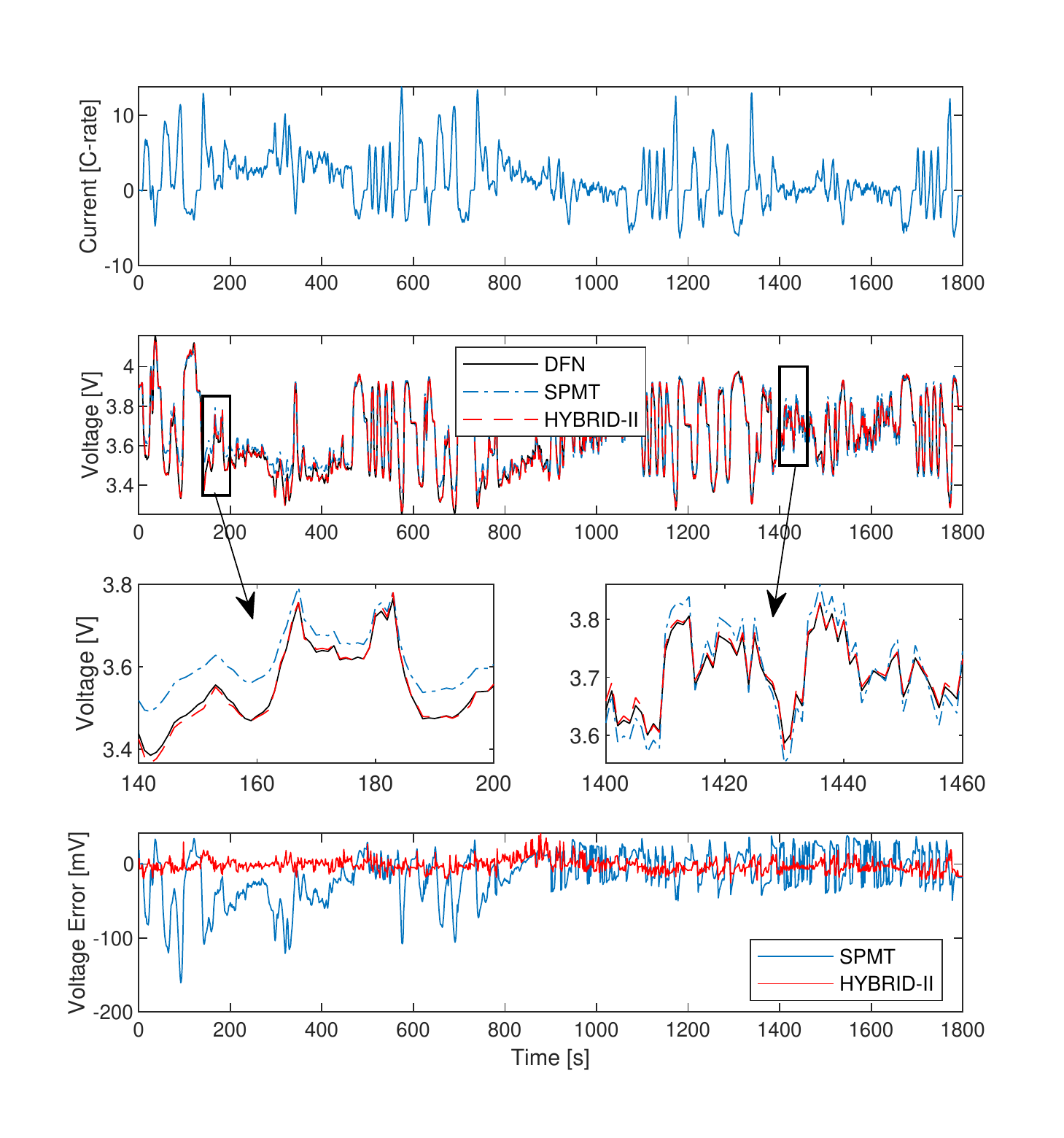}
    \caption{Testing results of HYBRID-\Rmnum{2} under US06 discharging for $\mathrm{SOC}(0) = 0.70$.}
    \label{HYBRID-II-test-US06}
\end{figure}

For the HYBRID-II model, its training performance is illustrated in Table~\ref{HYBRID-II-training-all-datasets}. Table~\ref{HYBRID-II-test-all-datasets} further shows its performance across the different test datasets, and Fig.~\ref{HYBRID-II-test-US06} displays its prediction under the US06-based test dataset. These results show that the HYBRID-II is also greatly effective in grasping and forecasting the dynamics of LiBs.


\begin{table}[h!]\centering
\ra{1.2}
\label{table}
\begin{tabular}{l | l l l}
\toprule
Input profile & RMSE (SPMT) & RMSE (HYBRID-\Rmnum{2}) & RER (\%) \\
\hline
CC-0.1C & 2.31 mV & 4.24 mV & -83.55\\

CC-0.2C & 4.37 mV & 3.43 mV & 21.51\\

CC-1C & 21.59 mV & 4.65 mV & 78.46\\

CC-2C & 56.40 mV & 4.57 mV & 91.90\\

CC-4C & 94.28 mV & 8.90 mV & 90.56\\

CC-6C & 118.86 mV & 8.75 mV & 92.63\\

CC-8C & 150.06 mV & 8.89 mV & 94.07\\

CC-10C & 191.05 mV & 13.03 mV & 93.18\\

UDDS-A & 23.32 mV & 8.48 mV & 63.64\\

US06 & 33.30 mV & 10.69 mV & 67.90\\
\bottomrule
\end{tabular}
\caption{Training results of the HYBRID-\Rmnum{2} under various current profiles and initial SOCs}
\label{HYBRID-II-training-all-datasets}
\end{table}

\begin{table}[h!]\centering
\ra{1.2}
\label{table}
\begin{tabular}{l | l l l}
\toprule
Input profile & RMSE (SPMT) & RMSE (HYBRID-\Rmnum{2}) & RER (\%) \\
\hline
CC-0.5C & 10.92 mV & 15.35 mV & -40.57 \\

CC-1C & 19.15 mV & 3.56 mV & 81.41 \\


CC-3C & 76.37 mV & 12.18 mV & 84.05\\


CC-5C & 101.85 mV & 7.80 mV & 92.34\\


CC-7C & 137.85 mV & 9.95 mV & 92.78\\


CC-10C & 207.29 mV & 11.98 mV & 94.22\\


UDDS-B & 17.59 mV & 8.37 mV & 52.42\\

US06 & 35.05 mV & 10.36 mV & 70.44\\
\bottomrule
\end{tabular}
\caption{Testing results of the HYBRID-\Rmnum{2} over the test datasets.}
\label{HYBRID-II-test-all-datasets}
\end{table}

Finally, we highlight that the HYBRID-I and HYBRID-II models provide higher testing accuracy and better predictive performance than the existing hybrid models for LiBs, as extensive simulation reveals. This underscores the efficacy of the proposed design that feeds a physics-based model's state information into the ML model.

\section{Conclusions}


The ever-increasing adoption of LiBs across various sectors presents a pressing demand for accurate and computationally efficient models. In this paper, we proposed two new hybrid models that integrate physics-based electrochemical modeling with data-driven ML to meet this need. The model development was driven by the novel perspective that informing the ML model of the electrochemical model's state information will bring significant improvements to the prediction performance. We conducted extensive simulations and illustrated that the proposed hybrid models can offer exceptionally high predictive accuracy for LiBs operating at a wide range of C-rates. This suggests their strong potential for enhancing various LiB   systems, especially when they run  at high power levels. Our future work will include experimental validation and the application of the proposed models to different battery management tasks.

\addtolength{\textheight}{-12cm}   
\balance
\bibliographystyle{unsrt}
\bibliography{ref}

\end{document}